%Paper: hep-th/9206038
%From: GOLDSTONE@irene.mit.edu
%Date: Tue, 9 Jun 1992 13:32:05 -0400 (EDT)

%%This is a generic, plain-TeX file.  No input macros are necessary.

\def\square{\kern1pt\vbox{\hrule height 1.2pt\hbox{\vrule width 1.2pt\hskip 3pt
   \vbox{\vskip 6pt}\hskip 3pt\vrule width 0.6pt}\hrule height 0.6pt}\kern1pt}
\hfuzz=30pt
\magnification=1200
\hoffset=-.1in
\voffset=-.2in

\vsize=7.5in
\hsize=5.6in
\tolerance 10000

\baselineskip 12pt plus 1pt minus 1pt
\pageno=0
\centerline{\bf A COMMENT ON THE RELATIONSHIP BETWEEN}
\smallskip
\centerline{{\bf DIFFERENTIAL AND DIMENSIONAL RENORMALIZATION}
\footnote{*}{This work is supported in part by funds
provided by the U. S. Department of Energy (D.O.E.) under contract
\#DE-AC02-76ER03069, and CICYT (Spain) under contract \#AEN90-0040.}}
\vskip 24pt
\centerline{Gerald Dunne}
\vskip 12pt
\centerline{\it Mathematics Department}
\centerline{and}
\centerline{\it Center for Theoretical Physics}
\centerline{\it Massachusetts Institute of Technology}
\centerline{\it Cambridge, Massachusetts\ \ 02139\ \ \ U.S.A.}
\vskip 12pt
\centerline{and}
\vskip 12pt
\centerline{Nuria Rius}
\vskip 12pt
\centerline{\it Center for Theoretical Physics}
\centerline{\it Laboratory for Nuclear Science}
\centerline{\it and Department of Physics}
\centerline{\it Massachusetts Institute of Technology}
\centerline{\it Cambridge, Massachusetts\ \ 02139\ \ \ U.S.A.}
\vskip 1.5in
\centerline{Submitted to: {\it Physics Letters B}}
\vfill
\centerline{ Typeset in $\TeX$ by Roger L. Gilson}
\vskip -12pt
\noindent CTP\#2100\hfill June 1992
\eject
\baselineskip 24pt plus 2pt minus 2pt
\centerline{\bf ABSTRACT}
\medskip
We show that there is a very simple relationship between differential and
dimensional renormalization of low-order Feynman graphs in renormalizable
massless quantum field theories.  The beauty of the differential approach is
that it achieves the same finite results as dimensional renormalization
without the need to modify the space time dimension.
\vfill
\eject
The method of ``differential regularization and renormalization'' was
introduced in Ref.~[1] by Freedman, Johnson and Latorre (FJL)
as a novel technique for regulating the ultraviolet
divergences of quantum field theories and computing finite
renormalized correlation functions.  The method is applied to real space
(rather than momentum space) Feynman diagrams, and requires no explicit
cut-off and no explicit counterterms.  Renormalized amplitudes are calculated
directly, and these satisfy Callan--Symanzik-type equations which determine
the renormalization group functions $\beta(g)$ and $\gamma(g)$ [1,2].  More
recent investigations have explored the compatibility with supersymmetry [3]
and gauge invariance [4], the extension to {\it massive\/} quantum field
theories [5], and the issue of lower-dimensional gauge theories [6].

In this Letter we explain how differential regularization and renormalization
is related to dimensional regularization and renormalization.  Such an
understanding is clearly important since dimensional regularization and
renormalization is the best understood and most widely-used procedure for
analyzing renormalizable
quantum field theories (particularly those involving {\it gauge\/}
symmetry).  Furthermore, a major motivation [1,2] for the development of
differential renormalization is the observation that (at least formally) the
method is compatible with both gauge and chiral symmetry, and may therefore
prove more convenient than dimensional renormalization for chiral gauge
theories as well as dimension-specific theories such as Chern--Simons,
Wess--Zumino--Witten, {\it etc\/}.

We shall illustrate the relationship between differential and dimensional
regularization and renormalization  in three
types of renormalizable {\it massless\/} quantum field theories:
\medskip
\item{(i)}massless scalar theory ($\phi^4$ in four dimensions, $\phi^3$ in six
dimensions, and $\phi^6$ in three dimensions);
\medskip
\item{(ii)}pure Yang--Mills theory in four-dimensions, using the background
field method; and
\medskip
\item{(iii)}massless four-dimensional quantum electrodynamics.
\medskip
\noindent We shall restrict our attention to low-order graphs in order to
pinpoint most clearly the {\it essential\/}
correspondence between differential and
dimensional renormalization (furthermore, much of the initial ``processing''
of higher-order graphs
is common to both approaches, and has been discussed in detail in
Refs.~[1,2]).  We shall also clarify the issue of gauge invariance in massless
QED, where differential renormalization introduces (in principle separate)
mass scales for the vertex and self-energy graphs which must however be
correlated in a specific way to satisfy the one-loop Ward identity.  This is
to be contrasted with the dimensional renormalization approach in which there
is a {\it single\/} mass scale $\mu$ (arising from the fact that the coupling
constant acquires a non-zero engineering dimension) and the Ward identity is
satisfied automatically (in, for example, the minimal subtraction scheme).

The essential idea [1] of differential regularization and renormalization
stems from the observation that most primitively divergent low-order
Feynman graphs are well-defined
in real space except for short-distance singularities at coincident points.
If these singularities are too severe, the graph will not have a well-defined
Fourier transform and higher-order graphs of which the original graph is a
sub-graph will contain divergent integrals --- hence the need for some form of
regularization.  The strategy of differential regularization and
renormalization
is to  {\it isolate the coincident point
singularities and to reduce the degree of divergence of these singularities by
expressing
the singular terms as derivatives of less singular terms\/}.
In practice, this strategy is implemented in two distinct stages.

In the first stage {\it no dimensionful regulator is needed\/}.
In this initial stage one may use straightforward algebraic manipulations and
simple differential identities in order to isolate the coincident point
singularities and to extract derivatives so as to reduce the degree of
divergence.  In massless quantum field theories (in $d\not=2$) we typically
encounter power-law-type singularities in bare graphs, and these may have
their degree of divergence reduced by using the {\it identity\/}
$$\left|x\right|^{-p} = {\square \left| x\right|^{-p+2} \over \left( -
p+2\right) \left( d-p\right)}\ \ ,\eqno(1)$$
In general (and, as we shall see, in particular for renormalizable theories)
one confront the singularity $|x|^{-d}$, whose degree of divergence may only
be reduced further by the introduction of an arbitrary (but essential)
logarithmic mass scale.\footnote{*}{Note also that $|x|^{-d}$ is the
borderline between powers of $|x|$ which have or do not have a well-defined
Fourier transform (see below, Eq.~(7)).}\footnote{$^\dagger$}{In a
{\it super-renormalizable\/}
theory this may not be necessary --- it may be possible to apply differential
renormalization without introducing a mass scale.  See Ref.~[6] for an
example.}  It is clear that one cannot use Eq.~(1) as it stands
because of the ${1\over d-p}$ pole.

In the second stage of differential
regularization and renormalization one proceeds by {\it defining\/} the
regulated form:
$$|x|^{-d}\bigg|_{\rm reg} \colon = {1\over 2(2-d)}\square \left( {\ln
M^2|x|^2 \over |x|^{d-2}}\right)\ \ ,\qquad (d\not=2)\ \ .\eqno(2)$$
This relation is certainly true away from the origin, and, furthermore, the
dependence on the arbitrary (but necessary) mass scale $M$ yields the
appropriate $\delta$-function singularity at the origin:
$$M{d\over dM} \left( |x|^{-d} \bigg|_{\rm reg}\right) = {2\pi^{d/2}\over
\Gamma\left( {\displaystyle{d\over 2}}\right)} \delta^{(d)} (x)\ \ .\eqno(3)$$
To obtain Eq.~(3) we have used the fact that the massless
scalar Green's function in $d$ ($\not=2$)
dimensions is
$$G(x) = - {\Gamma \left( {\displaystyle{d\over 2}}-1\right)\over 4\pi^{d/2}}
|x|^{2-d}\ \ ,\eqno(4)$$
and it satisfies
$$\square G(x) = \delta^{(d)} (x) \ \ .\eqno(5)$$
It is worth stressing that in differential regularization and renormalization
there are {\it no infinite\/} ({\it or finite\/}) {\it counterterms} --- one
may simply replace $|x|^{-d}$ by its regulated form as in (2).  Then the
scaling relation (3) is the key to deriving the Callan--Symanzik equations
satisfied by the renormalization amplitudes [1].  In the differential
approach, one also integrates by parts freely within graphs and in taking
Fourier transforms [1].  The consistency of this procedure and its
compatibility with unitarity has been explicitly checked (in massless $\phi^4$
theory through three-loop order) in a cutoff version of differential
renormalization [7].

This two-stage implementation of the differential renormalization strategy has
been applied with impressive success and efficiency to many examples of
renormalizable quantum field theories [1--7].  We now show that this same
strategy may be implemented using real-space {\it dimensional\/}
regularization and renormalization.  Recall that real-space dimensional
renormalization of quantum field theories, known
as the ``method of uniqueness,'' [8,9,10]
is an enormously powerful calculational tool
with which multi-loop computations may be performed with far greater ease than
with conventional (principally momentum space) techniques.  Here, however, our
emphasis and motivation are quite different --- we are interested in the
comparison with differential renormalization, and consequently we are
interested in not only the infinite parts of graphs, but also their regulated
finite parts.

The first important point to realize is that the {\it initial\/} stage of
``processing'' a bare Feynman graph (as outlined above) does not require a
regulating mass scale, and may be performed in {\it exactly\/} the same manner
in an arbitrary dimension $d$ of space-time. It is only in the second stage,
when one encounters the singularity $|x|^{-d}$, that any difference between
dimensional and differential regularization arises.  The dimensional
regularization approach described here is simply another way of regulating
this singularity.  In fact, if the theory is formulated in the non-integer
space-time dimension $d = D - 2\epsilon$ (where $D$ is the ``real'' integer
dimension and $\epsilon$ is infinitesimal) then the initial processing of the
bare graph does not lead to a singularity $|x|^{-d}$ but to a singularity
$|x|^{-d + r\epsilon}$, where $r\epsilon$ is some integer multiple of
$\epsilon$.  Indeed, since in a massless $d = (D - 2\epsilon)$-dimensional
theory the coupling constant acquires a mass dimension proportional to
$\epsilon$, one in fact encounters a singularity
$\mu^{r\epsilon}|x|^{-d+r\epsilon}$ where $\mu$ is a universal dimensionful
parameter associated with the coupling constant.\footnote{*}{The fact that the
``correct'' power of $\mu$ appears with the bare graph is a special feature of
renormalizable theories.}
Now one is free to use the identity (1) directly, isolating
the ${1\over d-p}$ pole as a $1/\epsilon$ pole:
$$\eqalign{\mu^{r\epsilon}|x|^{-d+r\epsilon} &= {1\over
\epsilon}\mu^{r\epsilon} {1\over r(2-d+r\epsilon)}\square |x|^{-d+r\epsilon +
2} \cr
&= -{1\over \epsilon} \ {4\pi^{d/2}\over r\left( 2 - d+r\epsilon\right) \Gamma
\left( {\displaystyle{d\over 2}}-1\right)} \delta^{(d)} (x) + {1\over
2(2-d)}\square \left( {\ln \mu^2 |x|^2\over |x|^{d-2}}\right) + {\cal
O}(\epsilon)\ \ .\cr}\eqno(6)$$
Thus, in the dimensional approach, the $x=0$ singularity of the left-hand side
of Eq.~(6) is regulated as
a ``counterterm'' $\delta^{(d)}(x)$, with both a $1/\epsilon$ pole
and a finite $\epsilon^0$ coefficient,
and a finite regulated $\epsilon^0$ term of
exactly the same form as the differential regularization expression (2) (with
the dimensional regularization mass scale $\mu$ identified with the
differential regularization mass scale $M$).  As a matter of terminology, we
shall refer to the $1/\epsilon$ pole counterterm as the ``infinite
counterterm,'' the $\epsilon^0$ counterterm as the ``finite counterterm,'' and
the $\epsilon^0$ regulated term involving $\mu$ as the ``non-counterterm finite
part.''
This is precisely analogous to the situation in conventional momentum space
dimensional regularization, where the divergences of divergent momentum
integrals are regulated as Laurent expansions in $\epsilon$.  Indeed, we may
equivalently present Eq.~(6) in momentum space form by Fourier transforming
the left-hand side:

$$\eqalign{\int d^dx\,e^{ik\cdot x} \mu^{r\epsilon}|x|^{-d+r\epsilon} &=
{2^{r\epsilon} \pi^{d/2} \Gamma\left( {\displaystyle{r\epsilon\over 2}}\right)
\over \Gamma\left( {\displaystyle{d-r\epsilon\over 2}}\right)} \left(
{\mu\over |k|}\right)^{r\epsilon} \cr
&= - {1\over \epsilon}\  {4\pi^{d/2}\over r\left( 2 - d + r\epsilon\right)
\Gamma \left( {\displaystyle{d\over 2}}-1\right)} \cr
&\quad + {\pi^{d/2}\over
\Gamma\left({\displaystyle{d\over 2}}\right)}
\left( - \gamma_E + \psi \left(
{d\over 2} - 1 \right) + \ln \left({2\mu^2\over |k|^2}\right)\right)+ {\cal
O}(\epsilon)\ \ .\cr}\eqno(7)$$

Here $\psi(z) = {d\over dz} \ln\Gamma(z)$ is the digamma function and
$\gamma_E=-\psi(1) = 0.5772\ldots$ is Euler's constant.  This agrees with the
Fourier transform of the right-hand side of Eq.~(6) computed by freely
integrating by parts (as is usual in dimensional regularization).  In fact,
using Eq.~(7) it is straightforward to convert {\it any\/} of the real space
expressions we discuss in this paper into the corresponding momentum space
formulas.

{\it Renormalization\/} now proceeds exactly as usual in the
dimensional regularization approach (albeit in real space rather than in
momentum space) --- {\it i.e.\/} by defining an appropriate subtraction
scheme (see {\it e.g.\/} Ref.~[11]).
Note from Eq.~(3) that the scaling dependence of the
``non-counterterm finite part'' produces contributions of the ``finite
counterterm'' form.  In more conventional language
this simply corresponds to adjusting the renormalization scale.

There are,of course, minor technical complications (see below) in gauge
theories involving some tensor algebra and Lie algebra manipulations and in
fermionic theories involving some Dirac matrix algebra, but the {\it
essence\/} of the correspondence between differential and dimensional
regularization (at low order) lies in Eqs.~(2), (3) and (6).
\goodbreak
\bigskip
\noindent{\bf 1.\quad Massless Scalar Theories}
\medskip
\nobreak
Consider the Euclidean space massless scalar theory with action
$$S=\int d^dx\left( {1\over 2} \phi\square \phi - {\lambda \over n!}
\phi^n\right) \eqno(8)$$
where $d\not=2$ and $n\not=2$.  The scalar Green's function is as in (4) and
(5) and so, apart from tadpole contributions,
the lowest order two-point function is (see Fig.~1)
$$\eqalign{\Gamma(x) &= {\lambda^2\over (n-1)!} \bigl( G(x)\bigr)^{n-1} \cr
&={\lambda^2\over (n-1)!}\left( - {\Gamma\left( {\displaystyle{d\over
2}}-1\right) \over 4\pi^{d/2}}\right)^{n-1} |x|^{(2-d)(n-1)}\ \
.\cr}\eqno(9)$$
For relevant values of $d$ and $n$, this graph is singular at the origin and
it may have its degree of
divergence reduced by applying Eq.~(1).  For the theory to be renormalizable
(at this order) this must lead to a $\square \delta(x)$ counterterm --- this
requirement relates the space-time dimension $d$ and the interaction power $n$
as:
$$(2-d) (2-n) = 4 \eqno(10)$$
(where we have used that $\square |x|^{2-d}\sim \delta^{(d)}(x)$).
Thus, for example, we have $\phi^4$ theory in four dimensions, $\phi^3$ in six
dimensions or $\phi^6$ in three dimensions.  In these cases, the two-point
function is proportional to $|x|^{-2-d}$.  Applying formula (1) once, we arrive
at a $|x|^{-d}$ singularity.  Then using (2) we may simply write down the
finite renormalized amplitude in the differential approach:
$$\Gamma(x)\bigg|_{\rm diff.\ ren.} =- {\lambda^2\over (n-1)!} \left( -
{\Gamma\left({\displaystyle{d\over 2}}-1\right) \over
4\pi^{d/2}}\right)^{n-1} {1\over 4d(d-2)} \square\ \square \left( {\ln M^2
|x|^2\over |x|^{d-2}} \right)\ \ .\eqno(11)$$
In the dimensional approach, with $d=D-2\epsilon$ (and the integer part, $D$,
of the space-time dimension satisfies (10)) the coupling constant $\lambda$ is
replaced by $\lambda\to \lambda_0 \mu^{(n-2)\epsilon}$, where $\lambda_0$ is
dimensionless and $\mu$ is a universal mass scale.\footnote{*}{Another way to
regulate this massless $\lambda\phi^n$ theory
is to keep $d$ at its integer value but
take $n$ to be non-integer, $n=N+\delta$ with $N$ an integer: this is the
``$\delta$-expansion'' of Bender and collaborators [12].}
  Then
$$\Gamma(x) = {\lambda^2_0 \mu^{2(n-2)\epsilon}\over (n-1)!} \left( - {\Gamma
\left( {\displaystyle{D\over 2}} - 1 - \epsilon\right) \over
4\pi^{D/2-\epsilon}}\right)^{n-1}
|x|^{-2-d+2(n-2)\epsilon}\ \ .\eqno(12)$$
Using the relation (6) and expanding in powers of $\epsilon$ we obtain
$$\eqalign{
&\Gamma(x)\bigg|_{\rm dim\ ren.} =
{\lambda^2_0\over (n-1)!} \left( -{\Gamma\left( {\displaystyle{D\over
2}}-1\right) \over 4\pi^{D/2}}\right)^{n-2}\cr
&\times {1\over 16D} \left(-{1\over\epsilon}+ (n-2) \psi
\left( {D\over 2}-1\right) - (n-2) \ln \pi - {4(2D^2-D-2)\over D(D-2)^2}
\right) \square\delta(x) \cr
&- {\lambda^2_0\over (n-1)!} \left( - {\Gamma\left( {\displaystyle{D\over
2}}-1\right) \over 4\pi^{D/2}}\right)^{n-1} {1\over 4D (D-2)} \square\ \square
\left( {\ln \mu^2 |x|^2\over |x|^{D-2}}\right) + {\cal O}(\epsilon)\ \
.\cr}\eqno(13)$$

Thus we see that the dimensional renormalization two-point function agrees with
the differential renormalization result if we choose our dimensional
renormalization scheme to be one of subtracting both the infinite and finite
counterterms, leaving just the non-counterterm finite part, with the scales
$M$ and $\mu$ identified.  And because of the scaling relation (3), any other
scheme which retains any part of the finite counterterm simply corresponds to
rescaling the arbitrary mass scale $M$.
\goodbreak
\bigskip
\noindent{\bf 2.\quad Four-dimensional Yang--Mills Theory in Background Field
Method}
\medskip
\nobreak
In Ref.~[1], Freedman, Johnson and Latorre (FJL) used differential
regularization and renormalization to compute the one-loop two-point function
in the background field method.  Recall [13], that due to the explicit
background field gauge invariance of the background field method, the
two-point function is sufficient to determine the renormalization group
$\beta$-function.  The one-loop contribution to the effective action from the
gluon and ghost loops in Fig.~2 is
(after performing the Lie algebra contractions)
$$\eqalign{\Omega_2(B)
&= {1\over 2} \left[\hbox{(2.a)} + \hbox{(2.b)}\right]\cr
&= - {g^2_0 \mu^{4-d}\over 2} C_A \int d^dx\, d^dy\, B^a_\mu (x) B^a_\nu (y)
\cr
&\times \biggl[ - {d\over 4}\partial_\mu\partial_\nu \left( G(x-y)^2\right) +
d\left( \partial_\mu G (x-y)\right) \left( \partial_\nu G(x-y)\right) \cr
&\qquad +
4\partial_\mu \partial_\nu \left( G (x-y)^2\right) - 4\delta_{\mu\nu}\square
\left( G(x-y)^2\right) \biggr] \ \ .\cr}\eqno(14)$$
Here $B^a_\mu$ are the background gauge field potentials,\footnote{*}{Note
that $B^a_\mu$ coincides with $A^a_\mu$ of Ref.~[13], while in Ref.~[1]
$B^a_\mu = gA^a_\mu$.} $C_A$ is the quadratic
Casimir in the adjoint representation, and all derivatives are with respect to
$x$.  $G(x-y)$ is the massless scalar Green's function as in Eq.~(4).  After
performing some simple tensor algebra in (14) we find the manifestly
transverse expression
$$\eqalign{
\Omega_2(B) &= - {g^2_0 \mu^{4-d}\over 2} C_A \left( - {\Gamma \left(
{\displaystyle{d\over 2}} - 1\right) \over 4\pi^{d/2}}\right)^2 {\left( 8 -
{\displaystyle{15\over 2}}d\right) \over \left( 2 - 2d\right)} \cr
&\times \int d^dx\, d^dy\, B^a_\mu(x) B^a_\nu (y) \left( \partial_\mu
\partial_\nu - \delta_{\mu\nu} \square \right) |x-y|^{4-2d}  \ \
.\cr}\eqno(15)$$
Now, setting $d=4-2\epsilon$ and then using the relation (6) once we obtain
$$\eqalign{\Omega_2 (B) &= - {g^2_0\over 2} \mu^{2\epsilon} C_A \left(
{\Gamma(1-\epsilon)\over 4\pi^{2-\epsilon}}\right)^2 {(-22+15\epsilon)\over
(4\epsilon-6)}
\int d^dx\, d^dy\, B^a_\mu(x) B^a_\nu(y) \left( \partial_\mu
\partial_\nu - \delta_{\mu\nu} \square \right) |x-y|^{-4+4\epsilon} \cr
&=- {11C_A\over 96} \  {g^2_0\over\pi^2} \left( {1\over\epsilon}+
{131\over 66} + \gamma_E +
\ln\pi\right) \int d^dx\, d^dy\, B^a_\mu(x) B^a_\nu(y) \left( \partial_\mu
\partial_\nu - \square \delta_{\mu\nu}\right) \delta(x-y) \cr
&+ {11C_A\over 24} \left( {g_0\over 4\pi^2}\right)^2 \int d^dx\, d^dy\,
B^a_\mu(x) B^a_\nu(y) \left( \partial_\mu \partial_\nu - \square
\delta_{\mu\nu}\right) \square \left( {\ln \mu^2 |x-y|^2\over |x-y|^2}\right)
+ {\cal O}(\epsilon)\ \ .\cr}\eqno(16)$$
The $1/\epsilon$ pole part of (16) (from which one may deduce the
one-loop $\beta$-function) agrees with the momentum space
dimensional regularization computation [13], and the non-counterterm finite
part (also from which one may deduce the one-loop $\beta$-function)
agrees with the differential
regularization result of FJL (see Eq.~(II.G.22) of Ref.~[1]), with $M=\mu$.
Once again, in
terms of {\it dimensional\/} renormalization, {\it differential\/}
renormalization consists of a subtraction scheme in which only the
non-counterterm finite part is retained as the renormalized amplitude.
\goodbreak
\bigskip
\noindent{\bf 3.\quad Massless Four-Dimensional QED}
\medskip
\nobreak
In $d$ dimensions, the massless fermion Green's function is
$$S(x) = - \left( \gamma_\mu {\partial\over \partial x^\mu}\right) G(x)
\eqno(17)$$
where $G(x)$ is the massless scalar Green's function in (4), and so $S(x)$
satisfies
$$\gamma\cdot {\partial\over\partial x} S(x) = \square G(x) = \delta^{(d)}(x)\
\
.\eqno(18)$$
Then the one-loop self-energy graph in $d=4-2\epsilon$ dimensions is (see
Fig.~3)
$$\Sigma(x) =- \left( e_o \mu^\epsilon\right)^2 \gamma_\mu S(x)
\gamma_\nu \delta_{\mu\nu}
G(x)\ \ .\eqno(19)$$
Using the Dirac matrix relation $\gamma_\mu \gamma_\nu \gamma_\mu =
(d-2)\gamma_\nu$, we find
$$\eqalign{\Sigma(x) &=e^2_0 \mu^{2\epsilon} (1-\epsilon) \left( {\Gamma
(1-\epsilon) \over 4\pi^{2-\epsilon}}\right)^2 \gamma\cdot
{\partial\over \partial
x} |x|^{-4+4\epsilon} \cr
&={e^2_0\over 16\pi^2} \left( {1\over \epsilon} + \left[
\gamma_E + 1+ \ln \pi\right]\right) \gamma\cdot
{\partial\over \partial x} \delta(x) - {e^2_0\over 64\pi^2} \gamma\cdot
{\partial\over \partial x} \square \left( {\ln \mu^2 |x|^2\over |x|^2}\right)
+ {\cal O}(\epsilon)\ \ ,\cr}\eqno(20)$$
where in the last step we have used the relation (6).  This agrees
with the standard  ({\it e.g.\/} Ref.~[11]) one-loop momentum
space dimensional regularization result,  and the non-counterterm finite piece
agrees with the differential renormalization result of FJL [1], with the
differential renormalization mass scale $M_\Sigma$ taken equal to $\mu$.

The analysis of the one-loop QED vertex function (see Fig.~4)
in real space requires some
preliminary algebraic and tensor manipulations in order to isolate the
short-distance singularities.  (The reason for this is simply that the vertex
effectively depends on two space-time coordinates,
rather than just one as in all the previous
examples.) However, these ``stage one'' manipulations
are exactly the same as in the differential regularization approach (see [1]),
except that we work in $d$ rather than four dimensions.
The vertex is (see Fig.~4)
$$\eqalign{V_j (x,y,z)
&= \left( e_0 \mu^{2-d/2}\right)^3 \gamma_\mu S(x-z) \gamma_j S(z-y)
\gamma_\nu \delta_{\mu\nu} G(x-y) \cr
&= e^3_0 \mu^{3\epsilon} \left( 2\gamma_b \gamma_j \gamma_a - 2\epsilon
\gamma_a \gamma_j \gamma_b\right) \left({\partial\over \partial x^a} G(x-z)
\right) \left( {\partial\over \partial z^b} G(z-y)\right) G(x-y) \ \
,\cr}\eqno(21)$$
where in the last line we have specialized to $d = 4-2\epsilon$ dimensions and
we have used the Dirac matrix identity: $\gamma_\mu \gamma_a \gamma_j \gamma_b
\gamma_\mu = 2\gamma_b \gamma_j \gamma_a - (4-d) \gamma_a \gamma_j \gamma_b$.

Using the notation of FJL, we write the vertex as
$$V_j (x,y,z) \equiv - e^3_0 \mu^{3\epsilon} \left( 2\gamma_b \gamma_j
\gamma_a - 2\epsilon \gamma_a \gamma_j \gamma_b\right) V_{ab} (x-z, y-z)\ \
,\eqno(22)$$
thereby defining
$$V_{ab} (u,v) \equiv \left( {\partial\over \partial u^a} G(u) \right) \left(
{\partial\over\partial v^b} G(v)\right) G(u-v)\ \ .$$
Exactly as in FJL, one then isolates the singularity in $V_{ab}$ within the
trace part by writing (for example)
$$\eqalign{
V_{ab}(u,v) &= {\partial\over\partial u^a} \left( G(u) \left( {\partial\over
\partial v^b} G(v)\right) G(u-v)\right) - {\partial\over \partial v^b} \left(
G(u) G(v) \left( {\partial\over \partial u^a} G(u-v)\right) \right) \cr
&\quad + G(u) G(v)
\left( {\partial\over \partial u^a} \  {\partial\over \partial
v^b} - {\delta^{ab}\over d} \  {\partial\over \partial u}\cdot  {\partial\over
\partial v}\right) G(u-v) \cr
&\quad +
{\delta^{ab}\over d} G(u)G(v) {\partial\over \partial u}\cdot{\partial\over
\partial v} G(u-v) \cr
&\equiv \widetilde{V}_{ab} (u,v) + {\delta^{ab}\over d} G(u) G(v)
{\partial\over \partial u}\cdot{\partial\over \partial v} G(u-v) \ \
.\cr}\eqno(23)$$
Note [1] that $\widetilde{V}_{ab} (u,v)$ has a finite Fourier transform in
four dimensions, so we only need to regulate the remainder, $V_{ab} -
\widetilde{V}_{ab}$.  Using the defining relation (5) for the scalar Green's
function we find
$$V_{ab} - \widetilde{V}_{ab} = - {\delta^{ab}\over (4-2\epsilon)} \left(
\Gamma(1-\epsilon) \over 4\pi^{2-\epsilon}\right)^2 \delta(u-v) |u|^{-4 +
4\epsilon} \ \ .\eqno(24)$$
In the {\it differential\/} approach one would have $\epsilon\equiv0$ and so
one
would proceed (see [1]) by using the regulated form of $|u|^{-4}$ given by
Eq.~(2).  In the {\it dimensional\/} regularization approach we expand
$|u|^{-4+4\epsilon}$ as in Eq.~(6) (with a factor $\mu^{2\epsilon}$
``borrowed'' from the full vertex expression in (22); remembering that the
vertex has an {\it overall\/} mass factor $\mu^\epsilon$ due to the mass
dimension of the coupling $e=e_0\mu^\epsilon$) to obtain
$$\eqalign{
V_{ab}(u,v) &= \widetilde{V}_{ab} (u,v) - {\delta^{ab} \mu^{-2\epsilon}\over
64\pi^2} \left( {1\over \epsilon} + \left[ \gamma_E + {5\over 2} + \ln
\pi\right] \right) \delta(u) \delta(v) \cr
&\phantom{= \widetilde{V}_{ab} (u,v)\ } + {\delta^{ab} \mu^{-2\epsilon}\over
256
\pi^4} \delta(u-v) \square \left( {\ln \mu^2 |u|^2\over |u|^2}\right) \cr
&= - {1\over 64\pi^6} V^{\rm FJL}_{ab} (u,v) \bigg|_{\rm reg} - {\delta^{ab}
\mu^{-2\epsilon}\over 64 \pi^2} \left( {1\over \epsilon} + \left[ \gamma_E +
{5\over 2} + \ln \pi\right]\right)\delta(u)\delta(v)\  \ .\cr}\eqno(25)$$
If we set $d=4$, then
$V^{\rm FJL}_{ab}(u,v)\big|_{\rm reg}$ is the differential regulated
form of $V_{ab}$ found by FJL (see Eq.~(II.C.9) of Ref.~[1]).  Inserting (25)
into the full vertex expression (22) we find
$$V_j (x,y,z) = V^{\rm FJL}_j (x,y,z)\bigg|_{\rm reg} + e_0 \mu^\epsilon
{e^2_0\over 16\pi^2} \left( {1\over \epsilon} + \left[ \gamma_E +
{1\over 2} + \ln \pi\right]\right) \gamma_j \delta(x-z) \delta(z-y)\ \
,\eqno(26)$$
where $V^{\rm FJL}_j\big|_{\rm reg}$ is the final differential regulated form
in [1], with the differential renormalization mass scale for the vertex,
$M_V$, taken equal to $\mu$.  (Notice that the $\left[ \gamma_E + {5\over 2} +
\ln\pi\right]$ $\epsilon^0$ coefficient
in (25) becomes $\left[ \gamma_E + {1\over
2} + \ln \pi\right]$ in (26) because of the contraction of the
${1\over\epsilon}$ pole in $V_{ab}$ with the $\epsilon \gamma_a \gamma_j
\gamma_b$ prefactor in the full vertex expression in (22).)
Once again we see that the differential regularization and
renormalization approach yields the non-counterterm finite part of the
dimensional regularization approach.

It is straightforward to check that the dimensional renormalization
vertex in (26) and self-energy in (20) satisfy the one-loop Ward identity
$${\partial\over \partial z^j} V_j(x,y,z) = \bigl( \delta(z-x) - \delta(z-y)
\bigr) \Sigma(x-y) \ \ .\eqno(27)$$
This is immediately clear for the ${1\over\epsilon}$ pole part of the
counterterms, but takes a little work for the other terms.  Indeed, at first
sight, comparing the $\left[ \gamma_E + {1\over 2} + \ln \pi\right]$ factor
in (26) with the ``corresponding'' $\left[ \gamma_E + 1 + \ln\pi\right]$
factor in (20), it looks as though there is a discrepancy in the finite part.
However, differentiating $\left. V^{\rm FJL}_j\right|_{\rm reg}$
produces an extra finite counterterm contribution which ``corrects'' matters.
In contrast, in the differential renormalization approach there are {\it no
counterterms at all\/} in $\Sigma$ or $V_j$, so this extra finite
``counterterm'' contribution to ${\partial\over\partial z_j}V_j$ must be
cancelled.  This may be achieved simply by rescaling the masses --- if $\ln
\left( M_\Sigma/M_V\right) = {1\over 4}$ the one-loop Ward identity is
satisfied [1].

The implication of this is simply that while dimensional regularization and
renormalization (for example, with a pole subtraction
renormalization prescription) automatically
respects gauge invariance ({\it i.e.\/} the Ward
identity is satisfied), the differential regularization and renormalization
procedure (which has {\it no counterterms\/} at all) will, in general,
require correlations between the mass scales appearing in various regulated
graphs in order to preserve gauge invariance.
This is not in the least surprising, as it is clear from Eqs.~(2) and (3)
that adjusting the differential renormalization mass scale in a renormalizable
theory produces finite counterterms.

To conclude, we have shown that there is a very simple relationship between
differential and dimensional renormalization in low order graphs.  It is not
unreasonable to expect that a systematic application of these ideas to higher
order graphs may yield a proof that differential renormalization works just as
well as dimensional renormalization to all orders.  However, the more
interesting task is to apply differential renormalization to theories in which
dimensional regularization is problematical.
\goodbreak
\bigskip
\centerline{\bf ACKNOWLEDGEMENTS}
\medskip
We are very grateful to D.~Freedman and K.~Johnson for helpful
discussions.  NR is indebted to the MEC (Spain) for a Fulbright Scholarship.
\vfill
\eject
\centerline{\bf REFERENCES}
\medskip
\item{1.}D. Freedman, K. Johnson and J. Latorre, {\it Nucl. Phys.\/} {\bf
B371}, 329 (1992).
\medskip
\item{2.}D. Freedman, MIT preprint CTP\#2020 (September 1991), to appear in
{\it Proceedings of the Stony Brook Conference on Strings and Symmetries\/}.
\medskip
\item{3.}P. Haagensen, {\it Mod. Phys. Lett.\/} {\bf A7}, 893 (1992).
\medskip
\item{4.}D. Freedman, G. Grignani, K. Johnson and N. Rius, MIT preprint
CTP\#1991 (March 1992), {\it Annals of Physics\/}, in press.
\medskip
\item{5.}P. Haagensen and J. Latorre, Barcelona preprint, UB-ECM-PF 92/5, to
appear in {\it Phys. Lett. B\/}.
\medskip
\item{6.}R. Mu\~noz--Tapia, Durham preprint, DUR-HEP~92 (March 1992).
\medskip
\item{7.}D. Freedman, K. Johnson, R. Mu\~noz-Tapia and X.
Vilas\'{\i}s-Cardona, MIT preprint CTP\#2099 (May 1992).
\medskip
\item{8.}K. Chetyrkin, A. Kataev and F. Tkachov, {\it Nucl. Phys.\/} {\bf
B174} (1980) 345; A. Vasil'ev, Yu. Pis'mak and J. Honkonen, {\it Theor. Math.
Phys.\/} {\bf 47}, 465 (1981).
\medskip
\item{9.}D. Kazakov, {\it Phys. Lett.\/} {\bf B133}, 406 (1983).
\medskip
\item{10.}A. Kotikov, {\it Phys. Lett.\/} {\bf B254}, 158 (1991);
J. Gracey, {\it Phys. Lett.\/} {\bf 277B}, 469 (1992).
\medskip
\item{11.}P. Ramond, {\it Field Theory: A Modern Primer\/} (Benjamin Cummings,
Reading, MA, 1981).
\medskip
\item{12.}C. Bender, K. Milton, M. Moshe, S. Pinsky and L. Simmons, Jr., {\it
Phys. Rev. Lett.\/} {\bf 58}, 2615 (1987); {\it Phys. Rev.\/} {\bf D37}, 1472
(1988).
\medskip
\item{13.}G.~'t~Hooft, {\it Nucl. Phys.\/} {\bf B62}, 444 (1973);
L. Abbott, {\it Nucl. Phys.\/} {\bf B185}, 189 (1981), and references
therein.
\vfill
\eject
\centerline{\bf FIGURE CAPTIONS}
\medskip
\item{Fig.~1:}Lowest order two-point function in a $\lambda\phi^n$ massless
scalar field theory.  There are ($n-1$) internal lines.
\medskip
\item{Fig.~2:}One-loop graphs involving two external background fields.  Wavy
lines indicate quantum gluons and dashed lines indicate ghosts.
\medskip
\item{Fig.~3:}One-loop self-energy graph in QED.  Straight lines indicate
fermion propagators and wavy lines indicate photon propagators.
\medskip
\item{Fig.~4:}One-loop vertex graph in QED.  The index $j$ is the space-time
index of the vertex.
\par
\vfill
\end